\documentclass[twocolumn]{aastex631}

\usepackage{amsmath}
\usepackage{multirow}
\usepackage{hyperref}

\begin{document}

\title{Carbon burning cannot explain puffy hypervelocity white dwarfs}

\author[0000-0001-6970-1014]{Natsuko Yamaguchi}
\affiliation{Department of Astronomy, California Institute of Technology, 1200 E. California Blvd, Pasadena, CA, 91125, USA}

\author[0000-0002-6871-1752]{Kareem El-Badry}
\affiliation{Department of Astronomy, California Institute of Technology, 1200 E. California Blvd, Pasadena, CA, 91125, USA}

\author[0000-0001-9195-7390]{Tin Long Sunny Wong}
\affiliation{Department of Physics, University of California, Santa Barbara, CA 93106, USA}

\author[0000-0002-9632-6106]{Ken J. Shen}
\affiliation{Department of Astronomy and Theoretical Astrophysics Center, University of California, Berkeley, CA, USA}

\begin{abstract}
Several hypervelocity white dwarfs (HVWDs) with space velocities of $\gtrsim 1000\,\mathrm{kms}^{-1}$ have recently been discovered. One possible origin of these stars is the dynamically-driven double-degenerate double-detonation (D6) scenario, in which an accreting sub-Chandrasekhar mass carbon-oxygen (CO) WD detonates as a SN Ia. In this scenario, the less massive WD may survive its companion's detonation and be ejected as a HVWD. Most of the observed HVWDs are hotter and puffier than normal WDs, perhaps due to their recent proximity to a SN. In this work, we test whether these properties can be explained by long-lived stable carbon (C) burning in the interiors of CO WD donors triggered by a SN shock. We model the long-term evolution of CO WDs following rapid energy injection using 1D models.  We find that stable C burning can be ignited in CO WDs with masses of $0.95 - 1.10\,M_{\odot}$ if SN energy penetrates sufficiently deeply. The resulting born-again stars settle on the C-burning main sequence while they convert their interiors from C and O to Ne and Mg, where they have temperatures and radii similar to some of the observed HVWDs. However, the timescale over which C-burning WDs remain inflated is $\lesssim 10^5\,$yr, which is at least an order of magnitude shorter than the kinematic ages of observed hot HVWDs. We conclude that observed HVWDs are unlikely to be inflated by C burning. The stellar evolution of observed HVWDs remains an open problem.
\end{abstract}


\section{Introduction} \label{sec:intro}

Type Ia supernovae (SNe Ia) are energetic transients thought to result from the thermonuclear explosion of white dwarfs (WDs) in binary systems. Despite playing an important role in both cosmology and astrophysics, the detailed physics of this process has been a topic of debate for many years (see \citealt{Liu2023RAA} and \citealt{Ruiter2025} for recent reviews). The canonical single-degenerate scenario involves a WD accreting mass from a non-degenerate companion until it reaches the Chandrasekhar limit \citep{Whelan1973ApJ}. Alternatively, in the double-degenerate scenario, both components are WDs. One such model that has gained popularity is the dynamically-driven double-degenerate double-detonation (D6) scenario. Here, a carbon-oxygen (CO) WD accretes material from a less massive WD which leads to a helium shell detonation that triggers a core carbon (C) detonation \citep{Guillochon2010ApJL, Fink2010A&A, Dan2011ApJ, Dan2012MNRAS, Pakmor2013ApJL}.

In the D6 scenario, there are two possible fates of the donor. First, it could also detonate due to the impact of the incoming ejecta from the explosion of its companion \citep[e.g.][]{Tanikawa2019ApJ, Pakmor2022MNRAS, Boos2024ApJ}. If it instead survives, it will be flung out of the orbit as a hypervelocity WD (HVWD) moving at roughly the Keplerian orbital velocity prior to the explosion \citep[e.g.][]{Bauer2021ApJL, Braudo2024, Shen2025ApJ}.

Recently, several HVWDs have been discovered using data from the Gaia mission which are proposed to originate from the D6 scenario \citep{Shen2018ApJ, El-Badry2023OJAp, Hollands2025MNRAS}. These have typical 3D space velocities of $2000\,\mathrm{kms}^{-1}$ and are both hotter and more inflated than typical WDs, with radii of $\sim 0.03-0.2\,R_{\odot}$. They span a wide range of temperatures, from $\sim 8000$\,K to $\sim100,000\,$K. One object (D6-2) has been traced back to a SN Ia remnant, and the UV spectrum of another object (J0927-6335) has revealed likely evidence of pollution from a SN Ia ejecta \citep{Werner2024A&A}. 

If these objects are indeed surviving donors from the D6 scenario, their hot and puffy nature might be interpreted as a result of the recent impact of a SN shock. However, it has proven challenging to reconcile the observed properties and inferred ages of all the observed objects with theory. For example, \citet{Bhat2025A&A} modeled the long-term evolution of donors powered by shock heating of their outer layers and nickel decay, but found this could not inflate the WDs over sufficiently long timescales. Meanwhile, \citet{Shen2025ApJ} found that the cooler HVWDs may be explained by the cooling of very low-mass fully convective stars. However, this would require the donors to have experienced copious amounts of mass loss pre-explosion, and it is unclear whether this model can explain the hottest objects, where relatively massive progenitors must become fully convective. In addition, the cooling timescales of the WD post-genitors of Type Iax SNe studied by \citet{Zhang2019ApJ} are too short at the high masses implied from the observed velocities of the D6 HVWDs. Given such challenges, \citet{Glanz2024arXiv} proposed an alternative formation scenario involving a merger of two hybrid helium-carbon-oxygen WDs. 

In this work, we run 1D models to test whether core C burning ignited in CO WDs could explain the observed properties of HVWDs. This possibility is motivated by the works of \citet{Paczynski1972AcA} and \citet{Paczynski1973A&A} who modeled a linear series of pure carbon stars and found branches of thermally stable stars burning carbon. We build on these early works by using CO WD models with realistic compositions and exploring the relevant range of energies for SN shocks to study C burning stars in the context of the D6 scenario.  

This paper is structured as follows. In Section \ref{sec:method}, we describe how we construct our initial CO WD models and inject energies into their interiors. In Section \ref{sec:results}, we describe the evolution of these models, and in particular, whether or not they successfully ignite C. For those that do, in Section \ref{sec:hvwds_comparison}, we compare the WD properties predicted by our models to those of the observed HVWDs. Finally, in Section \ref{sec:conclusion}, we summarize our findings. 

\section{Methods} \label{sec:method}

We use the 1D stellar evolution code Modules for Experiments in Stellar Astrophysics (MESA, version r24.03.1; \citealt{Paxton2011, Paxton2013, Paxton2015, Paxton2018, Paxton2019, Jermyn2023ApJS}) to produce CO WD models, inject energy into their interiors, and study their long-term evolution. 

\subsection{CO WD models} \label{ssec:co_wd_models}

Given the velocities of the observed HVWDs and pre-explosion Roche geometry of the D6 scenario, \citet{Bauer2021ApJL} and \cite{El-Badry2023OJAp} estimate masses between $\sim0.8-1.0\,M_{\odot}$ (albeit with large uncertainties, and with the exception of D6-2, which has a lower inferred mass of $\sim 0.3\,M_{\odot}$ and may have a helium core). Therefore, we begin by creating CO WD models within this mass range.   

\begin{figure*}
    \centering
    \includegraphics[width=0.95\linewidth]{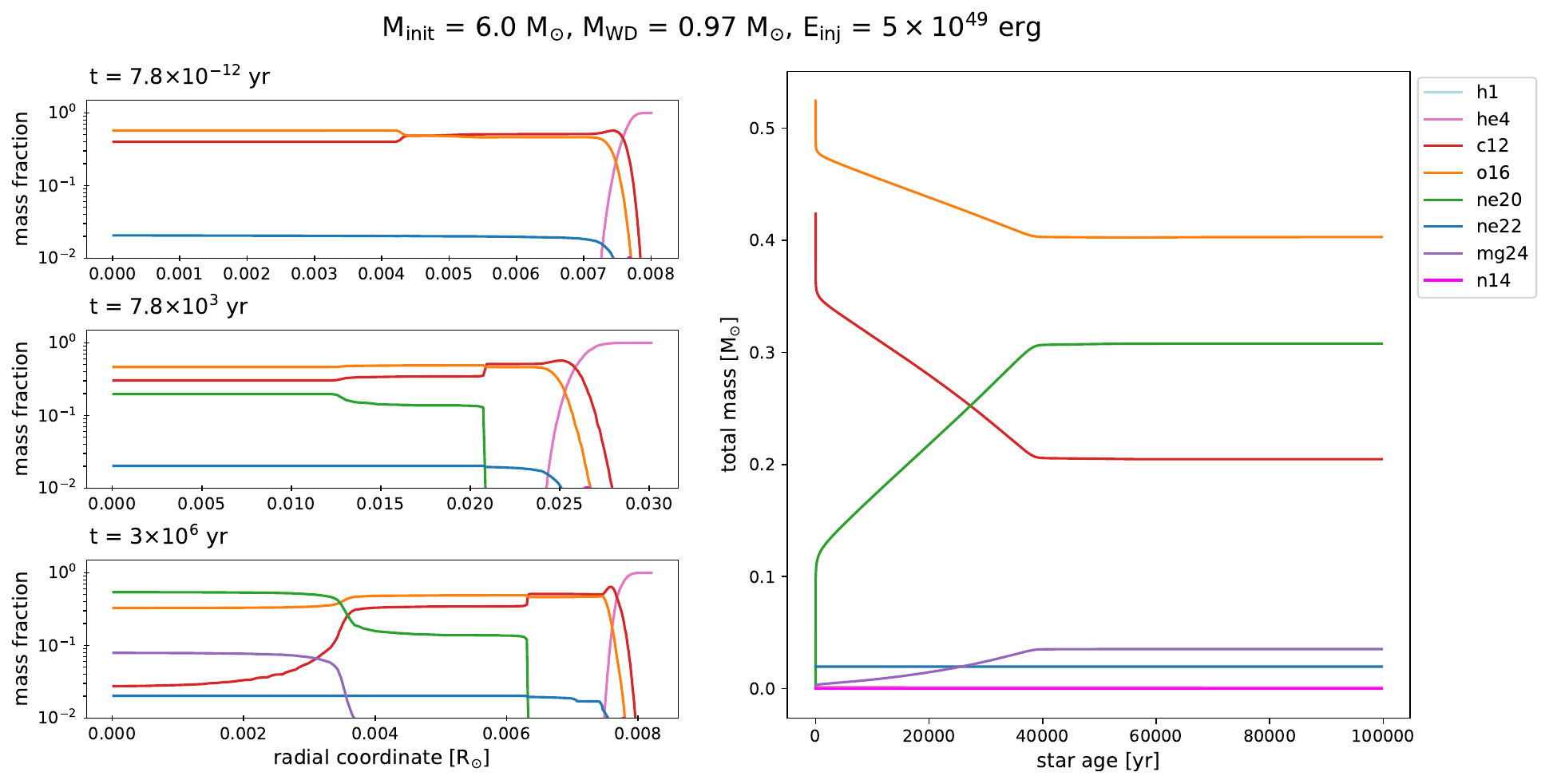}
    \caption{\textit{Left}: Abundance profiles in radial coordinates of the same model at different times. From top to bottom, we have the initial CO WD model, during C burning, and towards end of the run. 
    \textit{Right}: Temporal evolution of the total mass of several elements for the $0.98\,M_{\odot}$ WD model with $E_{\rm inj} = 5\times10^{49}\,$erg. Stable core burning occurs during a $\sim 4\times10^4$ year period in which $^{12}$C and $^{16}$O burn to $^{20}$Ne and $^{24}$Mg. By the end of this period, about 35\% of the WD's total mass has burned, including about 12\% that burns during an initial ``C flash".}
    \label{fig:6p0Msun_abun_profile}
\end{figure*}

We use the \texttt{make\_co\_wd} test suite, which evolves a star until the start of the thermally pulsating asymptotic giant phase and then strips off the outer envelope to leave behind a CO core.
For our fiducial models, we assume solar metallicity ($Z = 0.02$) and consider initial stellar masses of $M_{\rm init} = 4-6\,M_{\odot}$. This produces final CO WD masses of $M_{\rm WD} \sim 0.8-0.98\,M_{\odot}$. The top left panel in Figure \ref{fig:6p0Msun_abun_profile} shows the initial abundance profile for the $0.98\,M_{\odot}$ WD model. Models are terminated when their luminosity falls below $0.01\,L_{\odot}$. The nuclear reaction network used by the test suite, which includes carbon and oxygen burning, is used for all of our models (\texttt{co\_burn\_extras}, with a few special rate factors which can be found in the publicly available inlists). 

The donors in the D6 scenario may have formed via multiple episodes of interaction and mass transfer \citep[e.g.][]{Ruiter2009ApJ} and therefore could be more massive than the maximum mass of a CO WD produced by single-star evolution ($\sim 1.05\,M_{\odot}$, \citealt{Siess2007A&A}). Thus, we also construct $1.05$ and $1.1\,M_{\odot}$ CO WD models by letting the $0.98\,M_{\odot}$ model accrete a mixture of 65\% $^{16}$O and 35\% $^{12}$C by mass (roughly the initial core composition) using the \texttt{mass\_change} option. Nuclear burning is turned off during the accretion to speed up the computation and is turned back on afterwards. 

\subsection{Energy injection} \label{ssec:energy_inj_model}

We ignite the core of the model WDs by injecting a specified amount of total energy $E_{\rm inj}$, over a time interval, $\Delta t$. For our fiducial model, we inject energy such that the power injected per unit mass throughout the WD interior is $(E_{\rm inj}/\Delta t)/M_{\rm WD}$. In Appendix \ref{appendix:energy_inj_profile}, we test the effect of varying the energy injection profile such that it decreases in the innermost region of the WD. \citet{Bhat2025A&A} found that the shock induced by the SN explosion travels through the WD at $\sim 10^4\,\mathrm{kms}^{-1}$ and injected heat into their WD model over $\Delta t = 0.01$\,s. We find that some of our models fail to converge with such small values of $\Delta t$, so we instead use larger values, with a default of $\Delta t = 1$\,s. In Appendix \ref{appendix:Delta_T_tests}, we show that the value of $\Delta t$ has no effect on the long-term evolution of the WD, even for unphysically long values up to $\Delta t = 10^3$\,yr. 

Over a wide range of pre-explosion donor-to-accretor mass ratios $\sim 0.2 - 0.9$, the ratio of the donor radius to the orbital separation is in the range of $\sim 0.2-0.4$ \citep[][]{Bauer2021ApJL}. As a result, the donor's cross-section subtends a few percent of the sky as seen from the accretor. Given typical explosion energies of SNe Ia of $\sim 10^{\rm 51}\,$erg \citep[e.g.][]{Nomoto1984ApJ}, we expect the energy intercepted by the donor to be on the order of $10^{49} - 10^{50}\,\mathrm{erg}$. Thus, we test $E_{\rm inj}$ between these values, increasing it until we encounter numerical problems (typically above $\sim 7\times10^{49}\,$erg). Models are terminated after $3\,$Myr. The MESA inlists used here can be found in a Zenodo repository\footnote{10.5281/zenodo.16879323}. 

We emphasize that uniformly heating the WD is not a realistic model of the effects of the companion's explosion. Our goal here is to ignite core C burning, and to investigate whether C burning WD models can match the properties of observed HVWDs. In reality, a SN shock from the companion may fail to penetrate deeply enough to ignite C burning, or it may detonate the donor WD though a second double detonation \citep[][]{Boos2024ApJ}. 3D simulations are required to investigate these outcomes; this work is focused on the effects of {\it successful} C ignition. 

The surface layers of donors from the D6 scenario will be polluted by heavy elements from the SN ejecta \citep[e.g.][]{Bhat2025A&A}. This will affect the opacity of the outer layers and may change the timescale over which the WD relaxes. In Appendix \ref{appendix:increased_opacity}, we test this possibility and find that there is minimal change to the long-term evolution in the case of C ignition. 

\section{Results} \label{sec:results}

\subsection{Fiducial models} \label{ssec:results_fiducial}

\begin{figure*} 
    \centering
    \includegraphics[width=0.95\linewidth]{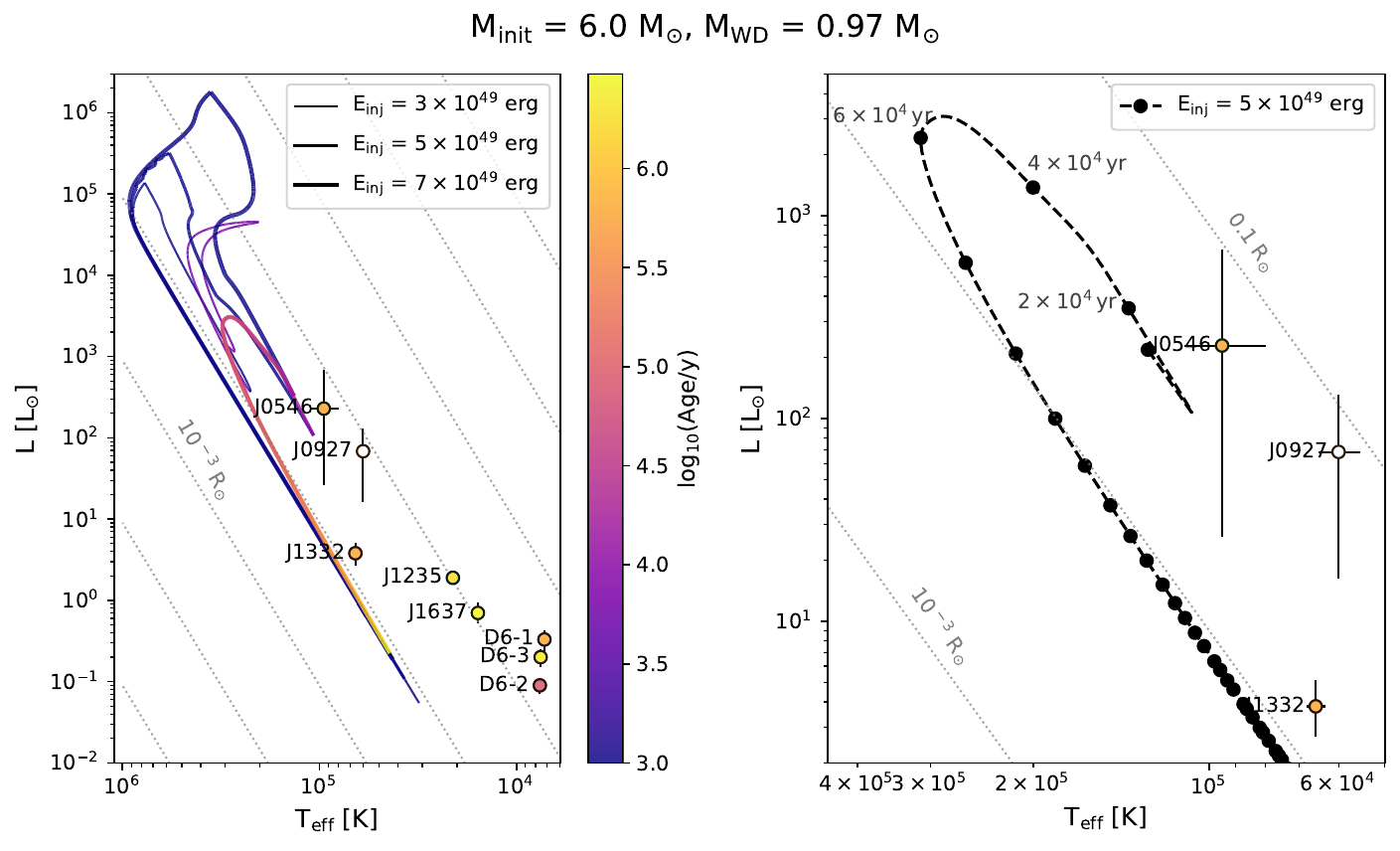}
    \caption{HR diagrams for the $0.98\,M_{\odot}$ WD models. We plot lines of constant radius for a blackbody, increasing from bottom left to top right in steps of a factor of 10. \textit{Left}: The tracks are colored by the stellar age and heavier line weights are used for those with larger $E_{\rm inj}$. The approximate locations of the 4 HVWDs from \citet{El-Badry2023OJAp} (with measurements from \citealt{Werner2024A&A} where available) and 3 from \citet{Shen2018ApJ} are plotted, with estimated temperatures and luminosities from \citet{Shen2025ApJ}. We also plot one HVWD originally identified by \citet{Raddi2019MNRAS}, with best-fit parameters from \citet{Hollands2025MNRAS}. These points are colored by ages inferred from flight times, with the exception of D6-2 which is associated to a SN remnant. The one unfilled point corresponds to J0927-6335 which is moving towards the Galactic midplane and therefore does not have an age constraint.\textit{Right}: Zoom-in on the $E_{\rm inj} = 5\times10^{49}\,$erg track after $10^3$ years. The markers show steps of $2\times10^4\,$yrs. We see that while the track approaches the locations of J0546 and J0927, this occurs after just $\sim 2\times10^4\,$yrs.}
    \label{fig:6p0Msun_HR}
\end{figure*}

\begin{figure*} 
    \centering
    \includegraphics[width=0.95\linewidth]{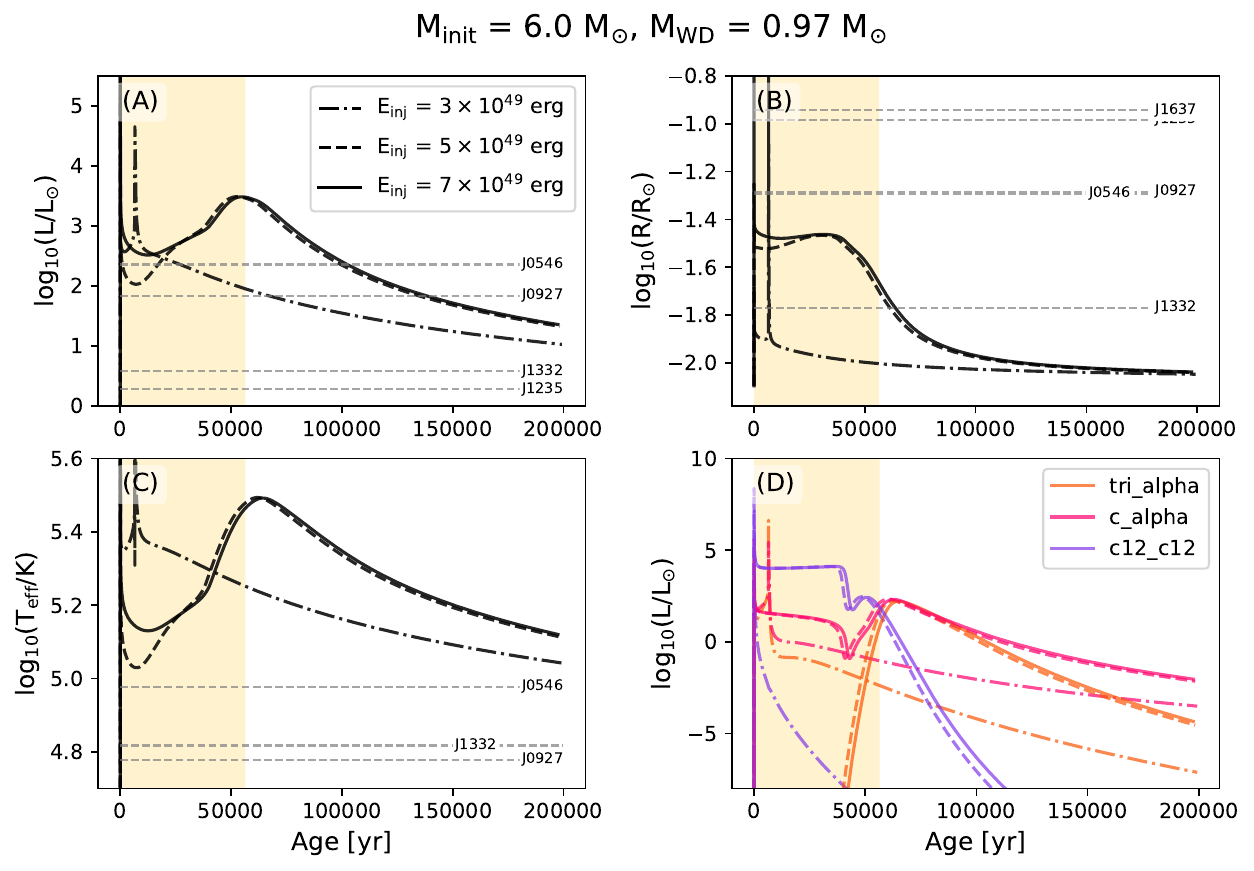}
    \caption{(\textit{A}), (\textit{B}), (\textit{C}): Time evolution of luminosity, radius, and effective temperature for the $0.98\,M_{\odot}$ WD models. The gray horizontal lines mark the values for the observed HVWDs. (\textit{D}): Time evolution of the energy production from several different nuclear reactions. The yellow shaded region highlights where C burning (\texttt{c12\_c12}) dominates the energy production for the $E_{\rm inj} = 5\times10^{49}$ and $7\times10^{49} \,$erg models, which lasts $\sim 6\times10^4\,$years.}
    \label{fig:6p0Msun_sps}
\end{figure*}

\begin{figure*}
    \centering
    \includegraphics[width=0.98\linewidth]{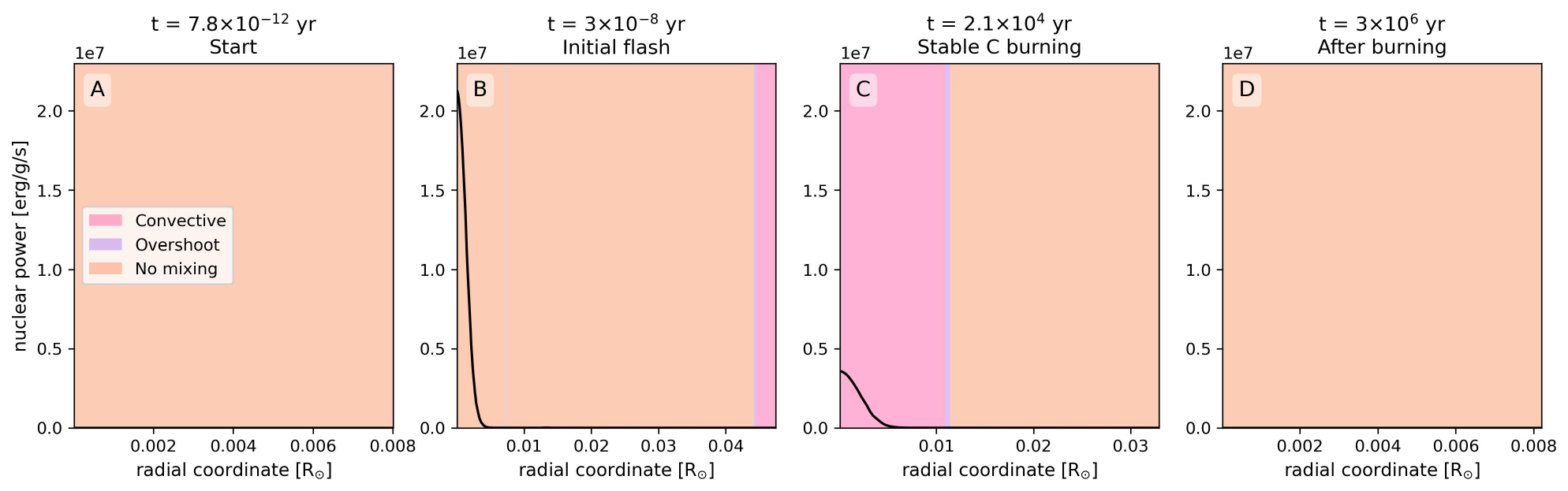}
    \caption{Radial profiles for the nuclear energy generation rate of the $0.98\,M_{\odot}, \, 5\times10^{49}\,$erg model at four different times. For comparison, all 4 panels have the same y-axis range. Regions with different mixing mechanisms are shaded with different colors. (\textit{A}): Initial profile. There is no mixing or burning throughout the WD. (\textit{B}): During the initial spike in burning. The peak at the core corresponds to the C burning which has been ignited. There is a thin convective region in the outemost layer. (\textit{C}): During stable C burning. The burning is confined to the central region ($\lesssim0.43\,M_{\odot}$) which becomes fully convective. (\textit{D}): Eventually, the burning terminates and the WD returns to its initial state.}
    \label{fig:6p0Msun_stellar_structure}
\end{figure*}

In Figure \ref{fig:6p0Msun_HR}, we show Hertzsprung-Russell (HR) diagrams for the $M_{\rm WD} = 0.98\,M_{\odot}$ models with varying $E_{\rm inj}$. And in Figure \ref{fig:6p0Msun_sps}, we plot the temporal evolution of several stellar parameters and nuclear energy generation rates. In both figures, we over-plot the properties of the 8 observed HVWDs thought to originate from the D6 scenario \citep{Shen2018ApJ, El-Badry2023OJAp, Hollands2025MNRAS}. 

We find that C burning is successfully ignited in models with $E_{\rm inj} = 5 \times10^{49}$ and $7\times10^{49}\,$erg. We see from panel D in Figure \ref{fig:6p0Msun_sps} that the energy production by C burning (\texttt{c12\_c12}; $\rm ^{12}C(^{12}C, \alpha)^{20}Ne$ and $\rm ^{12}C(^{12}C, p)^{23}Na$) remains roughly constant for the first $4\times10^4\,$yrs. It becomes subdominant to the other reactions after around $6\times10^4\,$yrs. Furthermore, the left panels in Figure \ref{fig:6p0Msun_abun_profile} show the abundance profiles for the $E_{\rm inj} = 5 \times10^{49}\,$erg model at different timesteps where we see the CO core being carved out. In the right panel, we show the total mass of different elements as a function of time. We find that about half of the total mass of carbon and $20\%$ of the oxygen initially present in the WD are burned, primarily into neon, by the end of the calculation. 

Figure \ref{fig:6p0Msun_stellar_structure} shows radial profiles of the nuclear energy generation rate for the $5\times10^{49}\,$erg model at different times. Regions of various mixing processes are shaded in distinct colors. From panel A, we see that the WD is initially convectively stable, and that there is virtually no burning\footnote{In cases where there is a small amount of hydrogen envelope left on the CO WD model, hydrogen burning processes can occur in the outer layers. However, this has a minimal impact on the subsequent evolution of the WD which is driven by core C burning.}. In panel B, we plot the profile during the initial spike in the nuclear burning, which lasts $\lesssim 100\,$yr. C burning (along with $\alpha$-capture reactions) is ignited in the still radiative core. There is a thin convective outer layer, likely a response to the initial energy injection. The WD is inflated, but at this point, this is primarily due to the expansion of its helium envelope outside the CO core. As energy is pumped into the WD, its surface luminosity exceeds the Eddington limit. Note that we do not turn on implicit hydrodynamics in MESA, and therefore during this phase, the star expands rapidly as opposed to driving out a wind. Such a wind may have a consequence on the mass loss and possibly the circumstellar medium \citep{Chandra2022MNRAS}. 

In panel C, we plot the profile once C burning becomes stable. During this time, burning is confined to the central region of the WD, which becomes fully convective, and the luminosity remains below $\sim 10\%$ of the Eddington limit. The WD core itself has now expanded, lifting its degeneracy. The nuclear burning luminosity at this stage significantly exceeds the surface luminosity, as much of the luminosity is lost to thermal neutrinos. Eventually, once all burning subsides, the WD returns to its initial radiative state (panel D).

We see that the $E_{\rm inj} = 5 \times10^{49}$ and $7\times10^{49}\,$erg tracks plotted in Figure \ref{fig:6p0Msun_sps} are almost identical. Thus, for cases where stable C burning proceeds, we find that the long-term evolution is relatively insensitive to the amount of energy initially injected. The WD radius returns to its initial value in $\sim 2\times10^5\,$yrs. Compared to the $E_{\rm inj} = 3 \times10^{49}\,$erg model, which does not persistently burn carbon, we see that the luminosity, radius, and temperature rise more gradually and peak at higher values. The brief burning episode at $\sim 7000\,$yrs in the $E_{\rm inj} = 3 \times10^{49}\,$erg model is dominated by $\alpha$-capture reactions and helium burning (i.e. $\texttt{tri\_alpha}$). This initially begins in the outer envelope, but convective mixing of helium pushes the burning deeper into the WD. C burning is never ignited. 

\begin{figure*}
    \centering
    \includegraphics[width=0.95\linewidth]{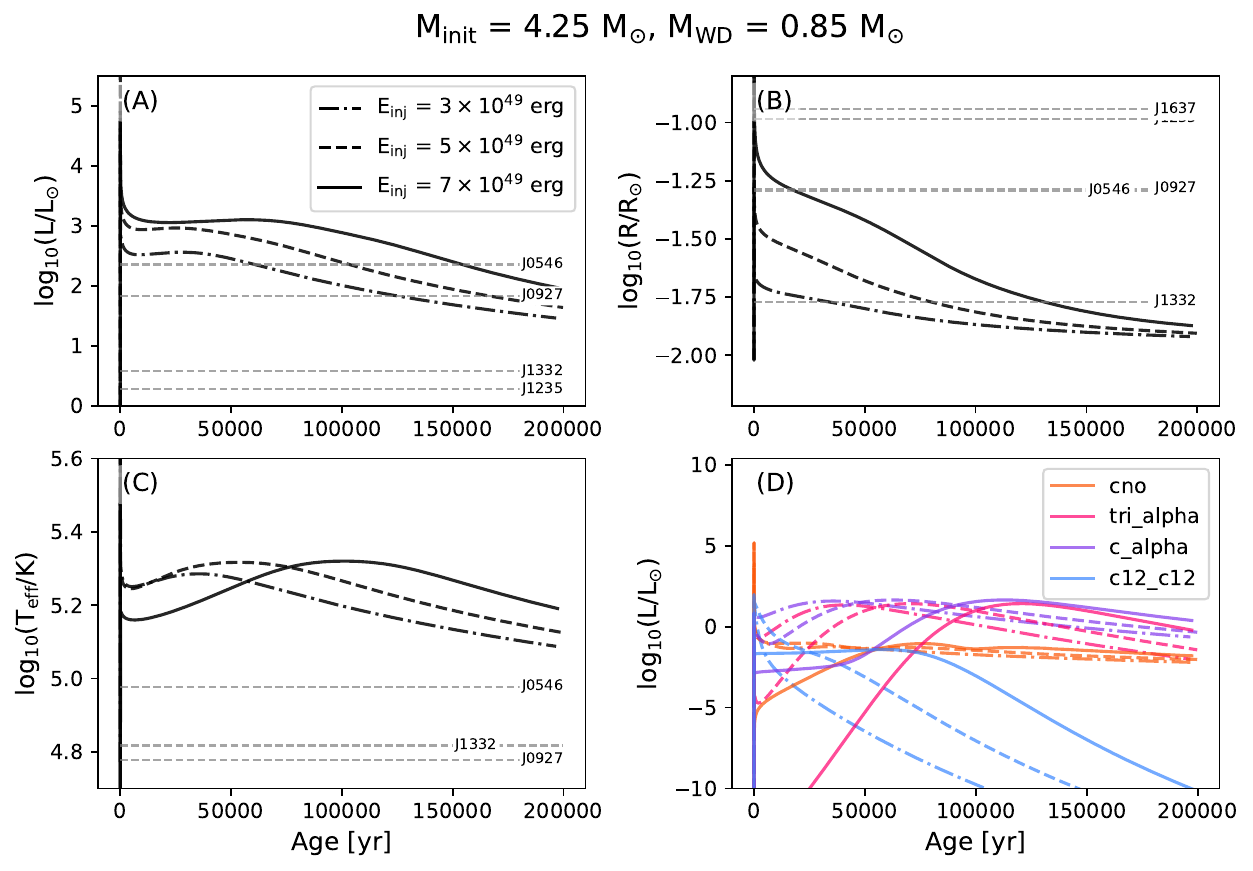}
    \caption{Analogous to Figure \ref{fig:6p0Msun_sps}, but for the  $0.85\,M_{\odot}$ WD models. In these models, there is no sustained C burning. The WD still expands as the energy deposited diffuses out of its interior.}
    \label{fig:4p25Msun_sps}
\end{figure*}

Figure \ref{fig:4p25Msun_sps} is analogous to Figure \ref{fig:6p0Msun_sps}  but shows the $0.85\,M_{\odot}$ WD models. For this mass, the luminosity from C burning is subdominant to the other reactions for all tracks and is several orders of magnitudes lower than for the $0.98\,M_{\odot}$ models described above (panel E). Additionally, we find that the total mass of carbon and oxygen remains nearly constant (to within $\sim 0.1\%$) throughout the run. We conclude that no significant C burning takes place in these models. Still, there is a temporary rise in luminosity, temperature, and radius of the WD as the energy injected diffuses out of the star. The timescale and magnitude of the rise are higher for larger amounts of energy injected. There is also substantial burning, primarily of hydrogen and helium, at early times. 

We obtained qualitatively similar outcomes for $0.8\,M_{\odot}$ WD models. As the WD evolution in these models is not driven by C burning, we do not study their properties further and instead point readers to \citet{Bhat2025A&A}, who performed a more detailed study of the evolution of WDs whose outer layers are shock heated by SN ejecta. Note that our models do not consider heating by the decay of $^{56}$Ni present in the outer layers of the WD from accretion of the ejecta, which \citet{Bhat2025A&A} found to be important in this case.  

For a $0.9\,M_{\odot}$ WD, we find that stable C burning occurs if $E_{\rm inj}$ is between $\sim 4-5\times10^{\rm 49}\,$erg. For larger values (e.g. $E_{\rm inj} = 7\times10^{49}\,$erg), after the first $\sim 10^{-8}\,$yr, the WD expands dramatically to over $10\,R_{\odot}$ and cools down to effective temperatures roughly an order of magnitude below the models with lower energies in the following year. Even after the WD undergoes Kelvin-Helmholtz contraction, the central temperature reached is lower by $\sim 150\,$million K. Accordingly, the power from nuclear burning falls below that of the lower $E_{\rm inj}$ models very rapidly, and no C burning is sustained. Therefore, a mass of about $0.9\,M_{\odot}$ may be understood to be a lower boundary for stable C burning to occur. 

\subsection{Massive CO WD models} \label{ssec:results_massive_wds}

\begin{figure*}
    \centering
    \includegraphics[width=0.95\linewidth]{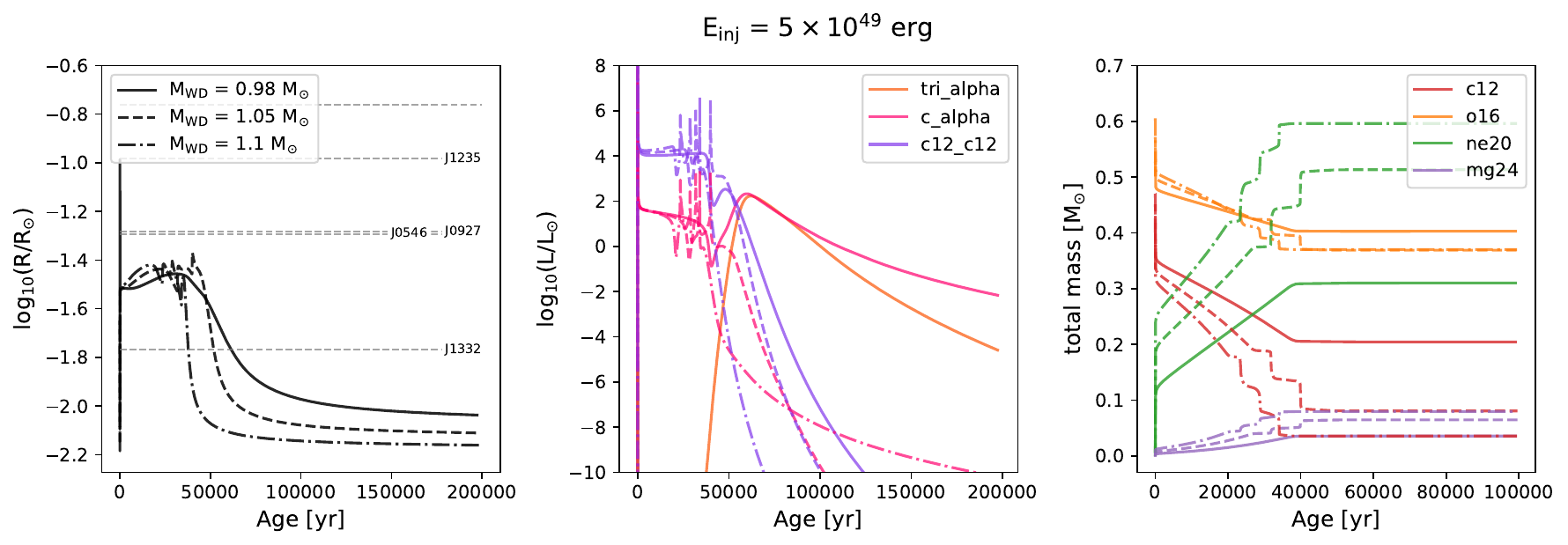}
    \caption{The first two panels are analogous to panels C and E of Figure \ref{fig:6p0Msun_sps}, but comparing the evolution for $0.98$, $1.05$, and $1.1\,M_{\odot}$ WD models with fixed $E_{\rm inj} = 5\times10^{49}\,$erg. The rightmost panel plots the temporal evolution of the total mass of elements for the same models, analogous to the plot in Figure \ref{fig:6p0Msun_abun_profile}. C burning is ignited in all three models but it lasts a shorter amount of time at higher masses.}
    \label{fig:massive_WDS_summary}
\end{figure*}

In Figure \ref{fig:massive_WDS_summary}, we compare the evolution of the $0.98\,M_{\odot}$ model with the $1.05$ and $1.1\,M_{\odot}$ models which were created via accretion of C/O-rich material (Section \ref{ssec:co_wd_models}), with $E_{\rm inj} = 5\times10^{49}\,$erg. All three models undergo a phase of C burning, but the period of burning, and thus the lifetime of the inflated WD, is shorter at higher masses. As shown in the rightmost panel, more massive WDs burn more of the available fuel, with the fraction of the total mass of carbon that burns increasing as $51\%$, $82\%$, and $92\%$ by the end of the run for the $0.98$, $1.05$ and $1.1\,M_{\odot}$ models. The final compositions of the latter two models are dominated by neon. There are  rapid fluctuations in the C burning power taking place between $\sim 3-5\times10^4\,$yr in the massive models. This feature is insensitive to changes in the spatial resolution. During this time, off-center C burning is ignited but is unable to propagate back to the center where most of the initial carbon has already been depleted. We see both regions of thermohaline convection and convective overshooting developing near the flame, which may also aid with the quenching \citep[e.g.][]{Siess2009A&A, Denissenkov2013ApJ}. This process is repeated several times, leading to the sharp spikes in burning luminosity. 

The trend in the duration of stable C burning with mass holds for lower mass models down to $0.9\,M_{\odot}$ created via single star evolution, meaning it is robust to the choices made in the engineering of the $1.05$ and $1.1\,M_{\odot}$ WDs. 

\section{Comparison to observed HVWDs} \label{sec:hvwds_comparison}

From the left panel of Figure \ref{fig:6p0Msun_HR}, we see that most of the observed HVWDs are more luminous at their temperatures than predicted by the models, reflecting their larger radii and cooler effective temperatures. This is similarly true for the more massive models described in Section \ref{ssec:results_massive_wds}. 
Furthermore, the observationally inferred inferred ages are $0.6-4.5$\,Myr (excluding D6-2), which range from a factor of a few to more than an order of magnitude longer than the $\sim 2 \times10^5\,$yr taken for the WD radii in our models to return to the WD cooling track. 

While J0546 and J1332 approach the tracks on the HR diagram, as seen on the right panel of Figure \ref{fig:6p0Msun_HR}, this occurs when the models are only $\sim 10^4\,$yr old, while the observed kinematic ages are $\sim 0.6$\,Myr. Although these kinematic age estimates are imprecise, an age of order $10^4$ years is very unlikely, given that the stars are not found within SN remnants. This tension persists across all models that burn C, even at lower masses down to $0.9\,M_{\odot}$. While it might be possible to further increase the energy injected, given that there is minimal difference between the $E_{\rm inj} = 5 \times10^{49}$ and $7\times10^{49}\,$erg models, it seems improbable for this to resolve the large discrepancy in timescales while remaining within a reasonable range of energies and preventing complete destruction of the WDs. 

For the linear series of pure C stars studied in \citet{Paczynski1972AcA}, thermally stable models with neutrino emission and total masses $\sim 0.8-0.9\,M_{\odot}$ have surface radii around $10^{-1.5}\,R_{\odot}$ and C burning luminosities on the order $10^{4}\,L_{\odot}$ (their Table 2, models 3 and 4). These values are comparable to our models during stable C burning (Figure \ref{fig:6p0Msun_sps}). Meanwhile, the nuclear timescales for the same \citet{Paczynski1972AcA} models are $\sim 10^{5.5}\,$yrs, which is about an order of magnitude longer than the $\sim 4\times10^{4}\,$yrs over which C burning is sustained in our models. However, for WD masses $\sim 0.9\,M_{\odot}$, our models do not burn all of their available fuel. Additionally, we are unable to get our WDs with masses below $0.9\,M_{\odot}$ to maintain burning. Differences in compositions, treatment of mixing processes, and thermal transport are likely a few factors that lead to the divergent outcomes. 

\section{Conclusion} \label{sec:conclusion}

In this work, we ran 1D models of the long-term evolution of CO WDs in which C burning is ignited following a companion's explosion. We summarize the key findings below: 
\begin{itemize}
    \item We used the 1D stellar evolution code MESA to evolve $4-6\,M_{\odot}$ stars to obtain CO WD models with masses $0.8 - 0.98\,M_{\odot}$. We also constructed massive $1.05$ and $1.1\,M_{\odot}$ WDs via accretion onto the $0.98\,M_{\odot}$ model (Section \ref{ssec:co_wd_models}). For our fiducial model, we uniformly deposited energy into the interiors of these models, varying the total amount of energy injected (Section \ref{ssec:energy_inj_model}). 
    \item  WDs with masses $\lesssim 0.85\,M_{\odot}$ do not undergo C burning after the initial energy injection (Figure \ref{fig:4p25Msun_sps}). 
    \item Given sufficiently large amounts of injected energy $\gtrsim 5\times10^{49}\,$erg (within the expected range of SN energy intercepted by the donor), WD models with masses $\gtrsim 0.9\,M_{\odot}$ can stably burn C for $\sim 5\times10^4\,$years, leading to them being significantly inflated for $\sim 10^5\,$years. Once C burning is successfully ignited, the evolution is relatively insensitive to the initial amount of energy injected (Section \ref{ssec:results_fiducial}, Figure \ref{fig:6p0Msun_sps}). 
    \item More massive WDs also undergo C burning, burning a larger fraction of their fuel, but the duration decreases with increasing mass (Section \ref{ssec:results_massive_wds}, Figure \ref{fig:massive_WDS_summary}).
    \item The majority of the observed HVWDs are too luminous at their temperatures to be consistent with any of the models. Furthermore, while the hottest HVWDs briefly approach the models on the HR diagram, their estimated ages of $\sim0.6\,$Myrs are over an order of magnitude longer than the time at which this occurs in the models (Section \ref{sec:hvwds_comparison}). This suggests that C burning is unlikely to explain the observed properties of these systems. 
\end{itemize}

\section{Acknowledgments}

We thank the anonymous referee for helpful comments that improved the manuscript.

This research benefited from discussions that were funded by the Gordon and Betty Moore Foundation through Grant GBMF5076.

This research was supported
by NSF grants AST-2508988 and AST-2307232 and by HST-GO-17441.001-A. 
NY acknowledges support from the Ezoe Memorial Recruit Foundation scholarship. TLSW acknowledges support from the Gordon and Betty Moore Foundation through grant GBMF5076. KJS is supported by NASA through the Astrophysics Theory Program (80NSSC20K0544) and by NASA/ESA Hubble Space Telescope program No.\ 17441.

\vspace{5mm}
\facilities{}

\software{MESA}

\appendix

\section{Changing the energy injection profile} \label{appendix:energy_inj_profile}

For our fiducial model, we took a fully uniform profile for the energy injection. Here, we test the effect of varying this profile so that less energy reaches the central region of the WD. We adopt a simple model where the energy per unit time per unit mass, $\epsilon(m)$, increases linearly up to a specified mass coordinate, $m = m_0$, after which it becomes uniform: 
\begin{equation} \label{eqn:epsilon}
\epsilon(m)=
\begin{cases}
    \epsilon_0f+\epsilon_0(1-f)\frac{m}{m_0}, & \text{if } m < m_0 \\
    \epsilon_0, & \text{if } m \geq m_0
\end{cases}
\end{equation} 
where $f$ is a free parameter that controls the steepness of the linear region, and $\epsilon_0$ is normalization factor which ensures that the equation integrates to the total energy injection rate, $E_{\rm inj}/\Delta t$.

\begin{figure*}
    \centering
    \includegraphics[width=0.9\linewidth]{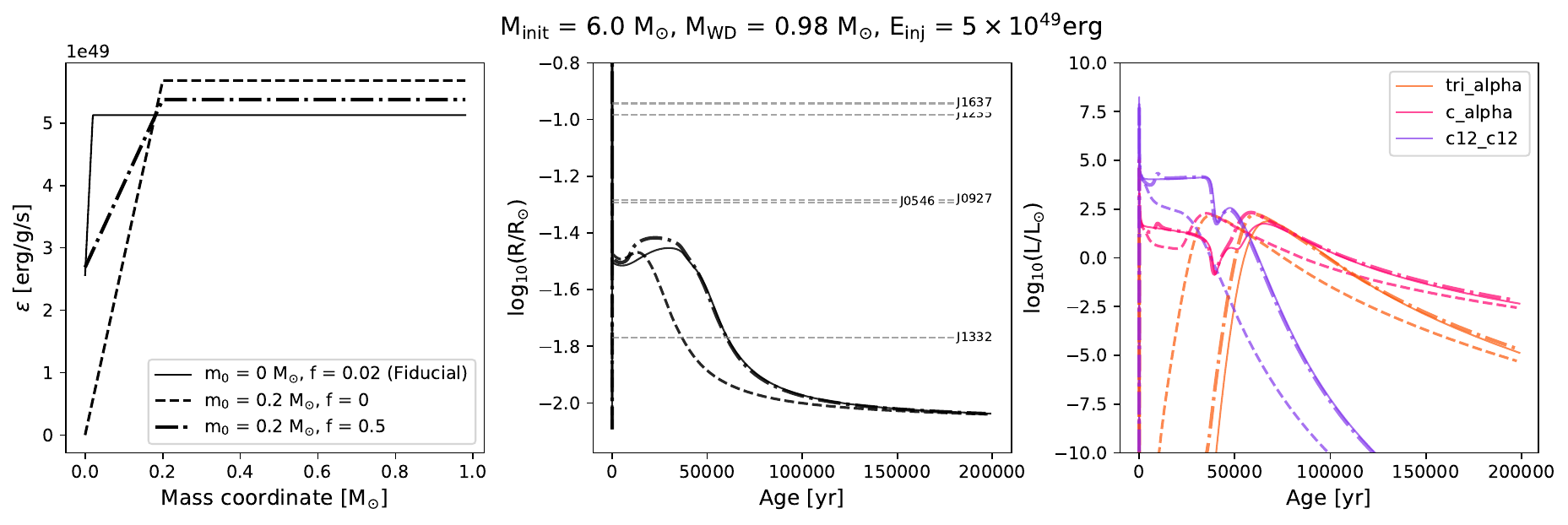}
    \caption{The leftmost panel shows the energy deposition profiles tested. The solid line is the fiducial model of a completely uniform profile ($m_0 = 0$). The right two panels are analogous to Figure \ref{fig:massive_WDS_summary}, but comparing the models with energy injected according to the different profiles.}
    \label{fig:epsilon_profiles}
\end{figure*}

In the leftmost panel of Figure \ref{fig:epsilon_profiles}, we plot the profiles tested. We include a linearly rising central region by setting $m_0 = 0.2\,M_{\odot}$ and vary its steepness by setting $f = 0, 0.5$. To the right, we plot the evolution of each of these models, analogous to Figure \ref{fig:massive_WDS_summary}. We find that for the $m_0 = 0.2\,M_{\odot}, f = 0$ model, the period during which C burning dominates the energy production is shorter and thus the WD deflates more quickly compared to the fiducial model. Indeed, this model only burns $\sim 20\%$ of the total mass of C. Meanwhile, if more energy reaches the core and there is a steeper transition to the uniform profile, as in the  $m_0 = 0.2\,M_{\odot}, f = 0.5$ model, the WD evolution is very similar to the fiducial case. 

Therefore, our fiducial assumption of a uniform profile provides us with a rough upper limit on the lifetime over which C burning dominates and the WD structure is impacted. Thus, changes to the profile do not change our conclusion that this process is inconsistent with the inferred ages of the observed HVWDs. 

\DeclareRobustCommand{\Deltat}{\texorpdfstring{$\Delta t$}{Δt}}

\section{Sensitivity to \Deltat } \label{appendix:Delta_T_tests}

\begin{figure*}
    \centering
    \includegraphics[width=0.8\linewidth]{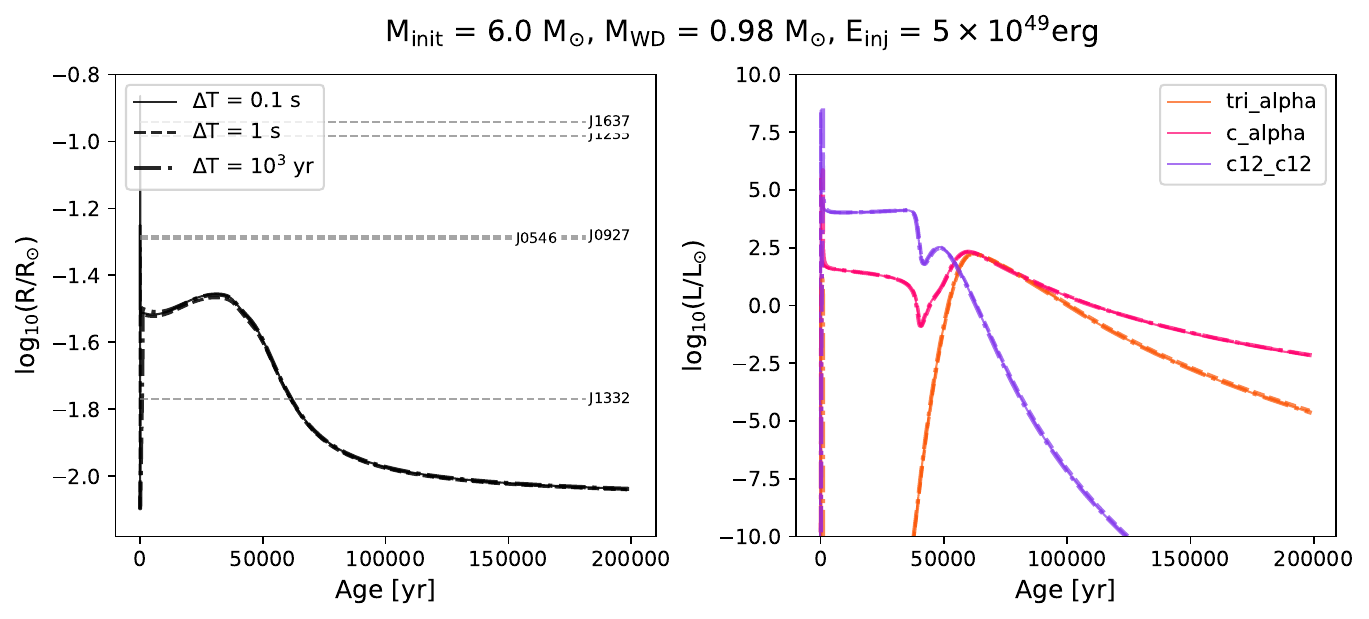}
    \caption{Analogous to Figure \ref{fig:massive_WDS_summary}, but varying $\Delta t$. We see that the evolution is indistinguishable for the $\Delta t=\rm0.1\,s$ and $\rm10^3\,yr$ models, which spans 11 orders of magnitude.}
    \label{fig:DeltaT_test}
\end{figure*}

In Figure \ref{fig:DeltaT_test}, we plot the time evolution of WD radii and luminosities from several nuclear reactions for $0.98\,M_{\odot}$ models with varying $\Delta t$ from $\rm 0.1\,s$ to $\rm10^3\,yr$ and fixed $E_{\rm inj} = 5\times10^{49}\,$erg. We find that models with $\Delta t \leq 10^3\,$yr all ignite C and the evolution of stellar properties are nearly identical. Their initial response to the energy injection in the first $\lesssim 1000\,$yr can differ slightly, with an expected delay in the peak of the C burning luminosity at longer $\Delta t$, but we see that this has almost no impact on the long-term evolution. However, above this, $\Delta t$ becomes comparable to the Kelvin-Helmholtz timescale of the WD and the energy is able to be transported out of the star efficiently, preventing sustained C burning. 

For some models, we find that if $\Delta t$ is too short, the solver struggles to converge. Based on the results of this test, in these cases, we allow $\Delta t$ to be increased. For all models presented in the rest of this paper, we keep it significantly below the maximum value tested, below $1\,$yr. Thus, our results are very likely insensitive to the precise value. 

\section{Increasing opacity} \label{appendix:increased_opacity}

\begin{figure*}
    \centering
    \includegraphics[width=0.8\linewidth]{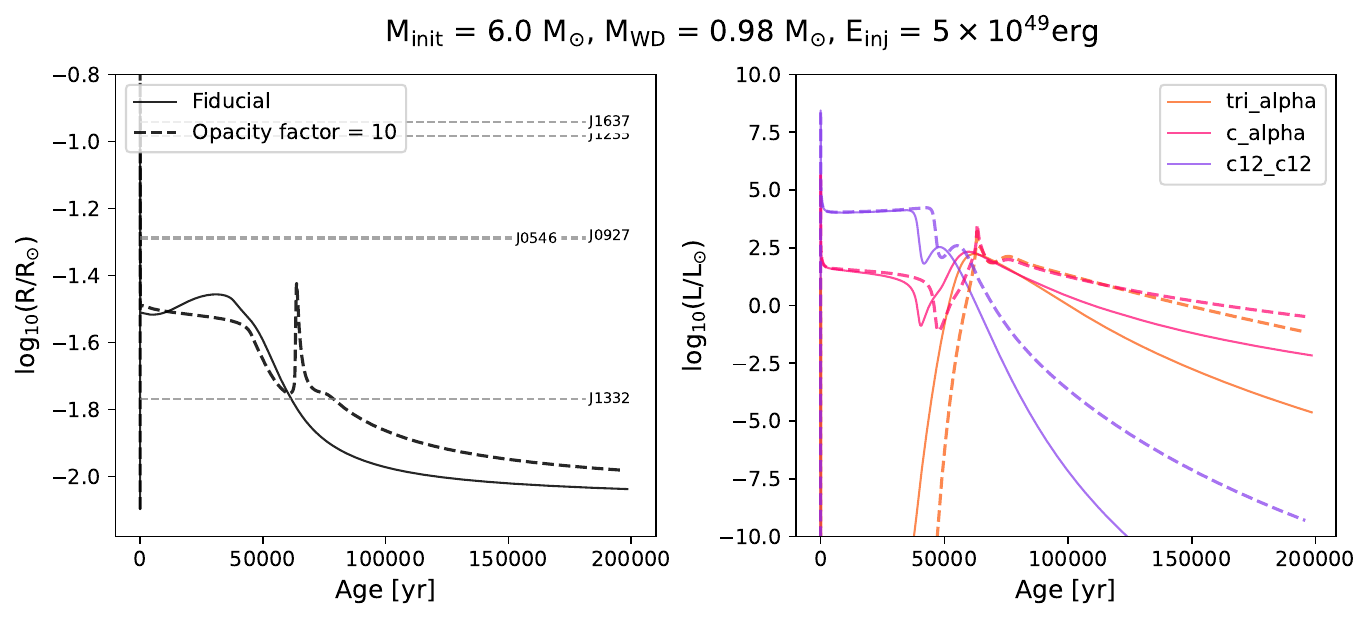}
    \caption{Analogous to Figure \ref{fig:massive_WDS_summary}, but comparing the fiducial model to one in which the surface opacity was increased by a factor of 10. }
    \label{fig:opacity_factor}
\end{figure*}

We test the effect of increased opacity resulting from the accretion of heavy elements by artificially increasing the opacity outside of the CO core by an order of magnitude using the \texttt{extra\_opacity\_factor} subroutine in MESA. The results are shown for the $0.98\,M_{\odot}$ and $E_{\rm inj} = 5\times10^{49}\,$erg model in Figure \ref{fig:opacity_factor}. We see that the lifetime of C burning, and thus the inflated WD, is slightly longer. However, the difference is insufficient to correct for the discrepancy with the observed systems. We note the small peak in radii (and luminosity) that occur at $\sim 6.3\times10^4\,$yr which corresponds to the peak in the power from the $\alpha$-capture of carbon (\texttt{c\_alpha}). 

\bibliographystyle{aasjournal}



\end{document}